\documentclass[prl,twocolumn,showpacs,amsmath,amssymb,superscriptaddress,floatfix]{revtex4}
\usepackage[english]{babel}
\usepackage{graphicx}
\begin{document}

\title{Entanglement of two delocalised electrons}

\author{A. Ram\v{s}ak}

\affiliation{Faculty of Mathematics and Physics, University of Ljubljana, Ljubljana,
Slovenia}

\affiliation{J. Stefan Institute, Ljubljana, Slovenia}

\author{I. Sega}

\affiliation{J. Stefan Institute, Ljubljana, Slovenia}

\author{J.H. Jefferson}

\affiliation{QinetiQ, St. Andrews Road, Great Malvern, England}

\date{20 March 2006}

\begin{abstract}
Several convenient formulae for the entanglement of two indistinguishable
delocalised spin-$\frac{1}{2}$ particles are introduced. This generalizes
the standard formula for concurrence, valid only in the limit of localised
or distinguishable particles. Several illustrative examples are given. 
\end{abstract}

\pacs{03.67.Mn, 03.67.Pp, 73.63.-b}

\maketitle
Entanglement is a well-defined quantity for two distinguishable qubits
in a nonfactorizable quantum state, where it may be uniquely defined
through von Neuman entropy and concurrence \cite{hill97,bennett96,vedral97,wooters98}.
However, amongst the realistic systems of major physical interest,
electron-qubits have the potential for a much richer variety of entanglement
measure choices due to both their charge and spin degrees of freedom.
For example, in lattice fermion models such as the Hubbard dimer,
entanglement is sensitive to the interplay between charge hopping
and the avoidance of double occupancy due to Hubbard repulsion, which
results in an effective Heisenberg interaction between adjacent spins
\cite{zanardi02}. In systems of identical particles the main challenge
is to define an appropriate entanglement measure which adequately
deals with multiple occupancy states 
\cite{schliemann01,ghirardi04,eckert02,gittings02,huang05,buscemi06}.
In the case of fermions such a measure must also account for the effect
of exchange \cite{vedral03} as well as of mutual electron repulsion.

Entangled fermionic qubits can be created with electron-hole pairs
in a Fermi sea \cite{beenakker05} and in the scattering of two distinguishable
particles \cite{bertoni03}. A spin-independent scheme for detecting
orbital entanglement of two-quasiparticle excitations of a mesoscopic
normal-superconductor system was also proposed recently \cite{samuelsson03}.

A consensus regarding the appropriate generalization of entanglement
measure which would consider spin and orbital entanglement of electrons
on the same footing has not, however, been reached yet. In any realistic
solid-state device, spin entanglement is intimately related to the
orbital degrees of freedom of the carriers, which cannot be ignored,
even in otherwise pure spin entanglement observations. In this paper
we introduce spin-entanglement measure formulae valid for real electrons
and show how, in general, spin-entanglement depends in an essential
way on spatially delocalised orbitals.

For two distinguishable particles $A$ and $B$, each described with
single spin-$\frac{1}{2}$ (or pseudo spin) states $s=\uparrow$ or
$\downarrow$ and in a pure state $|\Psi_{AB}\rangle=\sum_{ss'}\alpha_{ss'}|s\rangle_{\! A}|s'\rangle_{\! B}$
concurrence as a measure of entanglement is given by \cite{hill97}

\begin{eqnarray}
C & = & 2|\alpha_{\uparrow\!\uparrow}\alpha_{\downarrow\!\downarrow}-\alpha_{\uparrow\!\downarrow}\alpha_{\downarrow\!\uparrow}|.\label{eq:wooters}\end{eqnarray}
 Concurrence is related to the density matrix of a pair of spins \cite{wooters98}
and can be expressed in terms of spin-spin correlators $\langle\Psi_{AB}|S_{A}^{\lambda}S_{B}^{\mu}|\Psi_{AB}\rangle$
and expectation values $\langle\Psi_{AB}|S_{A(B)}^{\lambda}|\Psi_{AB}\rangle$,
where $S_{A(B)}^{\lambda}$ for $\lambda=x,y,z$ are spin operators
corresponding to spin $A$ or $B,$ respectively. This approach has
proved to be efficient in the analysis of entanglement in various
spin-chain systems with interaction \cite{osterloh02,syljuasen03,amico04,roscilde04}.

Consider now the general problem of two interacting electrons in a
pure state. It is clear that in some circumstances this system reduces
approximately to an equivalent system of two interacting spins, for
which the above entanglement formula is appropriate. Furthermore,
in the general case, entanglement between the spins of the fermions
relates to measurements of spin irrespective of their orbital motion.
We consider therefore spin-entanglement for a general class of two-electron
states on a lattice of the form

\begin{equation}
|\Psi\rangle=\sum_{i,j=1}^{N}[\psi_{ij}^{\uparrow\!\downarrow}c_{i\uparrow}^{\dagger}c_{j\downarrow}^{\dagger}+\frac{1}{2}(\psi_{ij}^{\uparrow\!\uparrow}c_{i\uparrow}^{\dagger}c_{j\uparrow}^{\dagger}+\psi_{ij}^{\downarrow\!\downarrow}c_{i\downarrow}^{\dagger}c_{j\downarrow}^{\dagger})]|0\rangle,\label{eq:psi}\end{equation}
 where $c_{is}^{\dagger}$creates an electron with spin $s$ on site
$i$ and $N$ is the total number of sites. The system in question
could be, for example, a tight-binding lattice containing two valence
electrons occupying non-degenerate atomic orbitals, or two electrons
in the conduction band of a semiconductor, for which the sites represent
finite-difference grid points. In either case, the interaction between
the electrons is included together with any externally applied potential.

\begin{figure}
\begin{center}\includegraphics[%
  width=60mm,
  keepaspectratio]{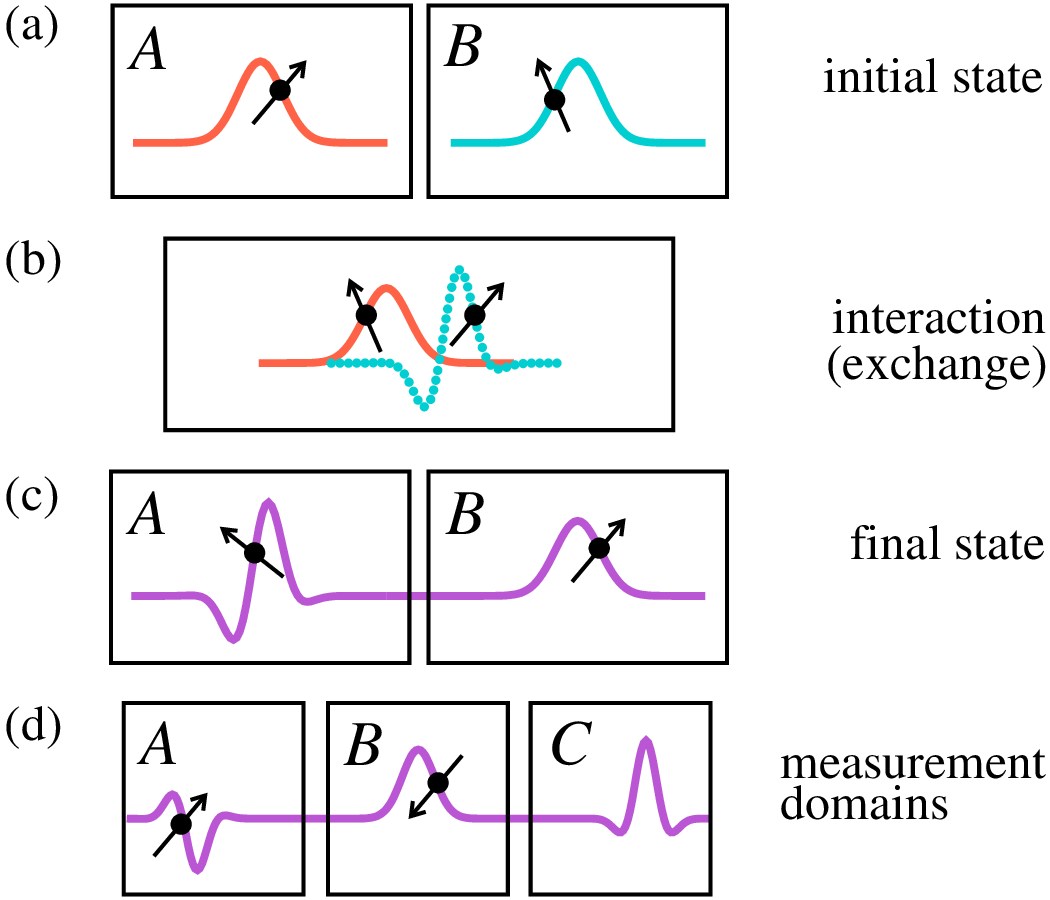}\end{center}

\caption{\label{cap:Fig1}(Color online) (a) In each of the domains $[A]$
and $[B]$ the probability of finding one electron is equal, $n_{A}=n_{B}=1$.
(b) Interacting electrons with possible exchange, (c) separated electrons,
and (d) several measurement domains, $n_{A}+n_{C}=n_{B}=1$.}
\end{figure}

The two electrons are in separate regions of space (measurement domains)
$[A]$ and $[B]$ as illustrated in Fig.~\ref{cap:Fig1}(a). Entanglement
might be produced, for example, when two initially unentangled electrons
in wave packets approach each other and interact {[}Fig.~\ref{cap:Fig1}(b){]}
and then again become well separated into distinct regions $[A]$
and $[B]$ {[}Fig.~\ref{cap:Fig1}(c){]}. Here one should realise
that in real measurements of entanglement, indistinguishable electrons
would be detected and the formalism relevant to distinguishable spins
is not directly applicable. Nevertheless, complete information regarding
the spin properties of such a fermionic system is contained in spin
correlation functions for the two domains. The spin-measuring apparatus
would measure spin correlation functions for two domains $[A]$ and
$[B]$ rather than for two distinguishable spins $A$ and $B$.

Concurrence as a measure of entanglement for two electrons is related
to the eigenvalues of the non-Hermitian matrix $\rho\tilde{\rho}$,
where $\rho$ is reduced density matrix given in terms of the electron
spin correlations corresponding to the domains, and $\tilde{\rho}$
is the time-reversed density matrix as in Ref.~\onlinecite{wooters98}.
In general the eigenvalues of $\rho\tilde{\rho}$ can be determined
only numerically and a closed form for concurrence can not be obtained,
unless the system exhibits additional symmetries. Possible symmetries
are conveniently studied through spin-spin correlation functions.
We express spin operators for domains $[A]$ and $[B]$ with fermionic
operators as the sum of operators for sites $i$ within the domain
$[A]$ (or $[B]$), i.e., $S_{A}^{\lambda}=\frac{1}{2}\sum_{i\in[A]}\sum_{ss'}c_{is}^{\dagger}\sigma_{ss'}^{\lambda}c_{is'}$,
where $\sigma^{\lambda}$ are Pauli matrices. 
For axially symmetric problems \cite{coulomb}, 
where $\langle\Psi|S_{A(B)}^{\lambda=x,y}|\Psi\rangle=0$
and $\langle\Psi|S_{A(B)}^{z}S_{B(A)}^{\lambda=x,y}|\Psi\rangle=0$,
concurrence may be written as
\begin{eqnarray}
C & = & \textrm{max}(0,C_{\uparrow\!\downarrow},C_{\parallel}),\label{eq:cmax}\\
C_{\uparrow\!\downarrow} & = & 2|\langle S_{A}^{+}S_{B}^{-}\rangle|-2\sqrt{\langle P_{A}^{\uparrow}P_{B}^{\uparrow}\rangle\langle P_{A}^{\downarrow}P_{B}^{\downarrow}\rangle},\nonumber \\
C_{\parallel} & = & 2|\langle S_{A}^{+}S_{B}^{+}\rangle|-2\sqrt{\langle P_{A}^{\uparrow}P_{B}^{\downarrow}\rangle\langle P_{A}^{\downarrow}P_{B}^{\uparrow}\rangle},\nonumber \end{eqnarray}
 where $S_{A(B)}^{+}=(S_{A(B)}^{-})^{\dagger}=\sum_{i\in A(B)}c_{i\uparrow}^{\dagger}c_{i\downarrow}$
are spin raising operators for domains $[A]$ or $[B]$ and $P_{A(B)}^{s}=\sum_{i\in A(B)}n_{is}(1-n_{i,-s})$,
with $n_{is}=c_{is}^{\dagger}c_{is}$, are spin-$s$ projectors operating
in domains $[A]$ (or $[B])$. Fermionic expectation values required
in Eq.~(\ref{eq:cmax}) are then given in terms of the amplitudes
in the normalised $|\Psi\rangle$ as\begin{eqnarray}
\langle S_{A}^{+}S_{B}^{-}\rangle & = & \sum_{[ij]}\psi_{ij}^{\uparrow\!\downarrow*}\psi_{ji}^{\uparrow\!\downarrow},\label{eq:sasb}\\
\langle S_{A}^{+}S_{B}^{+}\rangle & = & \sum_{[ij]}\psi_{ij}^{\uparrow\!\uparrow*}\psi_{ij}^{\downarrow\!\downarrow},\nonumber \\
\langle P_{A}^{\uparrow}P_{B}^{\downarrow}\rangle & = & \sum_{[ij]}|\psi_{ij}^{\uparrow\!\downarrow}|^{2},\nonumber \\
\langle P_{A}^{\downarrow}P_{B}^{\uparrow}\rangle & = & \sum_{[ij]}|\psi_{ji}^{\uparrow\!\downarrow}|^{2},\nonumber \end{eqnarray}
 where the summation in Eq.~(\ref{eq:sasb}) extends over all pairs
$[ij]$ such that $i\in[A]$ and $j\in[B]$. In analogy to the Bell
basis \cite{bennett96} one can introduce $\varphi_{ij}^{\pm}=(\psi_{ij}^{\uparrow\!\downarrow}\pm\psi_{ji}^{\uparrow\!\downarrow})/\sqrt{2}$
and $\chi_{ij}^{\pm}=(\psi_{ij}^{\uparrow\!\uparrow}\pm\psi_{ij}^{\downarrow\!\downarrow})/\sqrt{2}$
. $\varphi_{ij}^{\pm}$, e.g., are the amplitudes for creating two
electrons in a delocalised singlet or triplet state with zero total
spin projection. It then follows from Eqs.~(\ref{eq:cmax} and \ref{eq:sasb})
that the electrons are completely entangled, when either(i) $\varphi_{ij}^{-}=\imath c\varphi_{ij}^{+}$,
$\chi_{ij}^{\pm}=0$, or (ii) $\chi_{ij}^{-}=\imath c\chi_{ij}^{+}$,
$\varphi_{ij}^{\pm}=0$, where $c$ is a real constant. In the general
case (i.e., without spin symmetries) $C=1$ if $|\psi\rangle$ is
a linear combination of $AB$-entangled pair states, $|\psi\rangle=\sum_{[ij]}\psi_{ij}\sum_{\beta=1}^{4}b_{\beta}|ij,\beta\rangle,$
where $|ij,\beta\rangle$ are the Bell states \cite{bell} corresponding
to pairs $[ij]$ and $b_{\beta}$ are constants with $|\sum_{\beta=1}^{4}b_{\beta}^{2}|=\sum_{[ij]}|\psi_{ij}|^{2}=1.$

When $|\Psi\rangle$ is an eigenstate of the total spin projection
$S_{\textrm{tot}}^{z}$, Eqs.~(\ref{eq:cmax}) and (\ref{eq:sasb})
simplify further. In particular, $C=0$ if $S_{\textrm{tot}}^{z}=\pm1$,
while for $S_{\textrm{tot}}^{z}=0$ the concurrence is given solely
with the overlap between $|\Psi\rangle$ and the particular $AB$-spin-flipped
state $|\tilde{\Psi}\rangle=S_{A}^{+}S_{B}^{-}|\Psi\rangle$, Eq.~(\ref{eq:sasb}), or as,

\begin{eqnarray}
C & = & C_{\uparrow\!\downarrow}=|\sum_{[ij]}[(\varphi_{ij}^{+})^{2}-(\varphi_{ij}^{-})^{2}]|.\label{eq:cpsi}\end{eqnarray}
 If probabilities for singlet and triplet are equal, the concurrence
formula reduces to $C=2|\mathrm{Im\mathnormal{\sum_{\mathnormal{[ij]}}(\varphi_{ij}^{+})^{*}\varphi_{ij}^{-}|}}$
and if $\varphi_{ij}^{+}=\varphi_{ij}^{-}e^{\imath\delta}$, to $C=|\sin\delta|$.
If the state $|\Psi\rangle$ corresponds to the system in continuum
space, $i\rightarrow{\textbf{r}}=(x,y,z)$, the only change is that
summations are replaced by integrations of $\varphi^{\pm}=\langle{\textbf{r}}_{1},{\textbf{r}}_{2};S_{\textrm{tot}}|\Psi\rangle$
over the corresponding measurement domains, e.g., $C=|\int_{[A]}\!\int_{[B]}\![(\varphi^{+})^{2}-(\varphi^{-})^{2}]\textrm{d}^{3}{\textbf{r}}_{1}\textrm{d}^{3}{\textbf{r}}_{2}|$.

In order to illustrate how these concurrence formulae can be applied
in practice, as the first example we consider two interacting electrons
on a one-dimensional lattice with $N\rightarrow\infty$ and with the
hamiltonian, $H_{0}=-t_{0}\sum_{is}(c_{is}^{\dagger}c_{i+1,s}+h.c.)+\sum_{ijss'}U_{ij}n_{is}n_{js'}$.

To be specific, let one electron with spin $\uparrow$ be confined
initially to the region $A$ ($i\sim-L$) and the other electron in
region $B$ ($i\sim L)$ with opposite spin, Fig.~2(a). The simplest
initial state is two wave packets with vanishing momentum uncertainty
$\Delta k\rightarrow0$, the left with momentum $k>0$ and the right
with $q<0$. After collision the electrons move apart with probability
amplitude $t_{kq}$ for non-spin-flip scattering and spin-flip amplitude
$r_{kq}$. More general initial wave packets are defined with momentum
amplitudes $\phi_{k}$ and $\bar{\phi}_{q}$ for spin $\uparrow$
and $\downarrow$, respectively. Concurrence Eq.~(\ref{eq:cpsi})
after the collision is then expressed as

\begin{equation}
C=2|\int\!\!\!\int t_{kq}^{*}r_{kq}|\phi_{k}|^{2}|\bar{\phi_{q}}|^{2}\textrm{d}k\textrm{d}q|,\label{eq:ckq}\end{equation}
 which simplifies to $C\sim2|t_{kq}r_{kq}|$ for sharp momentum resolution
wave packets, with $k=-q=k_{0}$. Note that $C=1$ when spin-flip
and non-spin-flip amplitudes coincide in accord with recent analysis
of flying and static qubits entanglement \cite{jrr05,gunlycke05,giavaras06}
or of scattering of distinguishable particles \cite{bertoni03}.

Consider the prototype finite range interaction, $U_{ij}=\frac{1}{2}U\sum_{m=0}^{M}\delta_{|i-j|,m}$.
The Hubbard model ($M=0$) can be solved analytically in one-dimension
\cite{lieb} and the amplitudes are $t_{kq}=1+r_{kq}=(\sin k-\sin q)/[(\sin k-\sin q)+\imath U/(2t_{0})]$.
In Fig.~2(a) concurrence is presented for wave packets with well
defined momentum $k_{0}$ for $U=t_{0}$, together with a longer range
interaction case, $M=3$, for sharp momentum (full line) and for a
Gaussian initial amplitude $\bar{\phi_{k}}=\phi_{-k}$ with $\Delta k=\pi/10$
(dashed line). An interesting observation here is substantial reduction
of concurrence due to the coherent averaging in Eq.~(\ref{eq:ckq}).
Additionally, electrons will be completely entangled at some kinetic
energy comparable with the repulsion, $U\sim2t_{0}(1-\cos k_{0})$,
where spin-flip and non-spin-flip amplitudes coincide.%
\begin{figure}
\begin{center}\includegraphics[%
  width=60mm,
  keepaspectratio]{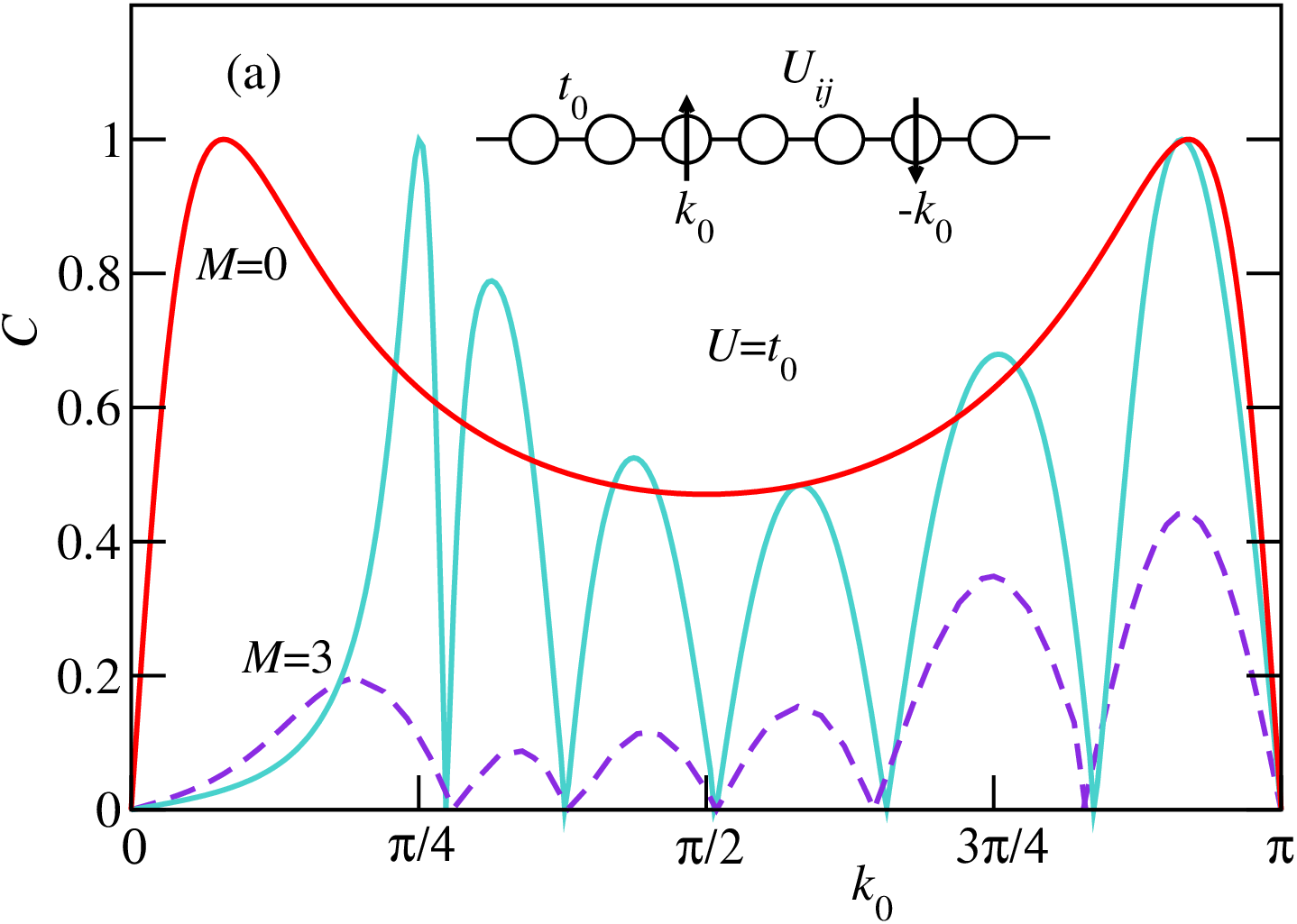}\end{center}

\begin{center}\includegraphics[%
  width=60mm,
  keepaspectratio]{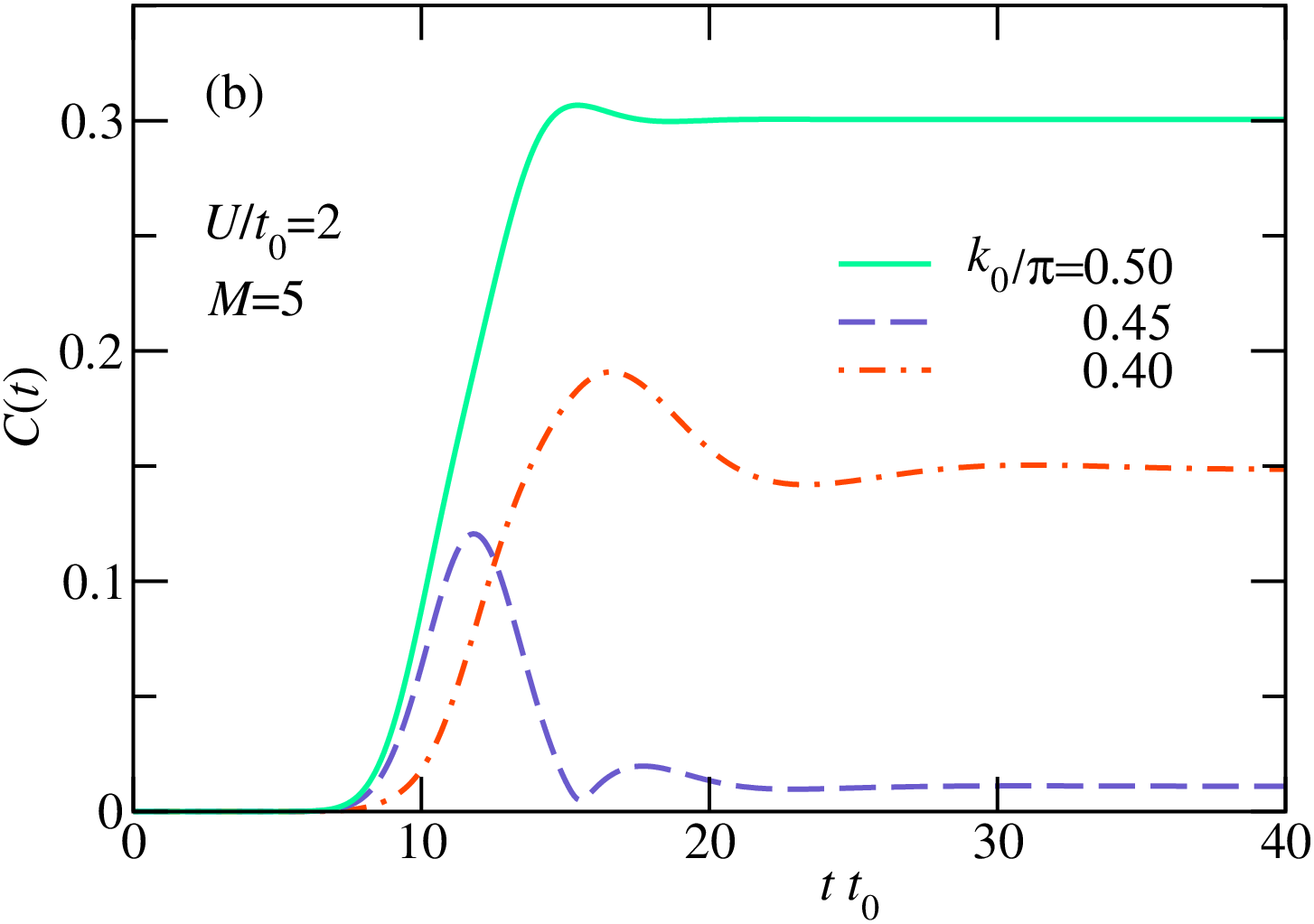}\end{center}

\caption{\label{cap:Fig2}(Color online) (a) $C$ for: (i) the Hubbard model
$(M=0)$ for $U=t_{0}$ and $\Delta k=0$; (ii) $M=3$: for $\Delta k=0$
(full line) and $\Delta k=\pi/10$ (dashed). (b) $C(t)$ for Gaussian
packets with various $k_{0}$ and $M=5$, $U=2t_{0}$ and $\Delta k=\pi/20$.
At $t=0$ the separation between the packets is $2L=10/\Delta k$.}
\end{figure}

The concurrence formula Eq.~(\ref{eq:cpsi}) is derived for electronic
states when double occupancy is negligible, i.e., $\psi_{ii}^{\uparrow\!\downarrow}\rightarrow0$,
which in our case is strictly fulfilled only asymptotically when the
electrons are far apart. However, Eq.~(\ref{eq:cpsi}) can be evaluated
at any time $t$ and the resulting $C(t)$ can serve as a measure
of entanglement during the transition from initial to final state.
In Fig.~2(b) we present the time dependence of $C(t)$ for some typical
$k_{0}$, with $M=5$ and $U=2t_{0}$. Oscillation with $t$ can be
interpreted as response to the finite time duration of electron-electron
interaction and the model can be approximately mapped onto an effective
Heisenberg model, for which concurrence oscillates as $C(t)=|\sin J_{\textrm{eff}}\, t|$,
where $J_{\textrm{eff}}$ is the effective antiferromagnetic coupling
between the electrons.

Another important example is the concurrence of flying--static qubits
in experiments in which the system is prepared with a static electron
bound in some confining potential (region $[B]$) and a flying electron
injected in some distant region $[A]$ \cite{jrr05,gunlycke05}. Contrary
to the previous case with translation symmetry, after the collision
there are nonvanishing amplitudes for transmission (into region $[C]$)
and reflection (back into region $[A]$), as shown in Fig.~1(d).

Let the initial state be prepared as $\varphi_{ij}^{\pm}=(b_{i}g_{j}\pm g_{i}b_{j})/\sqrt{2}$,
where $b_{i}$ is the orbital state of the bound electron with spin
$\downarrow$ centered around $i\sim0$. Similarly, $g_{j}\propto\int\phi_{k}e^{\imath[k(j+L)-\omega_{k}t]}\textrm{d}k$
is the initial orbital state of the propagating electron with spin
$\uparrow$, centered around $i\sim-L$ and moving in the positive
$i$-direction with momentum amplitude $\phi_{k}$ peaked at $k\sim k_{0}$,
and with momentum uncertainty $\Delta k\rightarrow0$. Here we consider
elastic scattering with amplitudes after the collision, $\varphi_{ij}^{\pm}=r_{\pm}(b_{i}a_{j}\pm a_{i}b_{j})+t_{\pm}(b_{i}c_{j}\pm c_{i}b_{j})$,
where $r_{\pm}(k_{0})$ and $t_{\pm}(k_{0})$ are singlet (triplet)
reflection and transmission amplitudes and $a_{j}$, $c_{j}$ are
normalised wave packets with mean momentum $-k_{0}$ and $k_{0}$,
respectively.

Two basic experimental setups are possible when electrons are detected
in different measurement domains\emph{,} $[AB]$ or $[BC]$. Concurrence
corresponding to reflected qubits is then \begin{equation}
C_{AB}=\frac{2|\langle S_{A}^{+}S_{B}^{-}\rangle|}{n_{A}n_{B}}\sim\frac{|r_{+}^{2}(k_{0})-r_{-}^{2}(k_{0})|}{|r_{+}(k_{0})|^{2}+|r_{-}(k_{0})|^{2}},\label{eq:ca}\end{equation}
 where $n_{A}=\langle\sum_{s,i\in[A]}n_{is}\rangle$, $n_{B}=1$ \cite{na}.
Concurrence for transmitted qubits, $C_{BC}$, is given by an analogous
expression with $A\rightarrow C$, and consequently with $r_{\pm}$
replaced with $t_{\pm}$. If the measuring apparatus captures both,
reflected and transmitted electrons ($i\in[A]\cup[C]$, $j\in[B]$),
concurrence is given by $C_{AC,B}=|(r_{+}-r_{-})^{*}(r_{+}+r_{-})+(t_{+}-t_{-})(t_{+}+t_{-})^{*}|$
and no additional renormalisation is required. Eq.~(\ref{eq:ca})
also follows directly from Eq.~(\ref{eq:wooters}) if appropriately
applied to scattering states \cite{jrr05,gunlycke05}. However, for
finite $\Delta k$, $C_{AB}$ (and correspondingly $C_{BC}$ or $C_{AC,B}$)
has to be rederived from Eq.~(\ref{eq:cpsi}),\begin{equation}
C_{AB}=\frac{|\int[r_{+}^{2}(k)-r_{-}^{2}(k)]|\phi_{k}|^{2}\textrm{d}k|}{\int[|r_{+}(k)|^{2}+|r_{-}(k)|^{2}]|\phi_{k}|^{2}\textrm{d}k}.\label{eq:cak}\end{equation}

In order to demonstrate the basic properties of $C_{AB}$ and $C_{BC}$
we consider here the Anderson model, $H=H_{0}+\sum_{s}[\epsilon n_{0s}-(t_{1}-t_{0})(c_{-1s}^{\dagger}c_{0s}+c_{0s}^{\dagger}c_{1s}+h.c.)]$,
where $H_{0}$ is the Hubbard hamiltonian in which $U=0$ except for
the impurity site, $\epsilon<0$ is the impurity energy level, and
$t_{1}$ is the hopping matrix element connecting the impurity site
$i=0$ with left and right leads.

\begin{figure}
\begin{center}\includegraphics[%
  width=65mm,
  keepaspectratio]{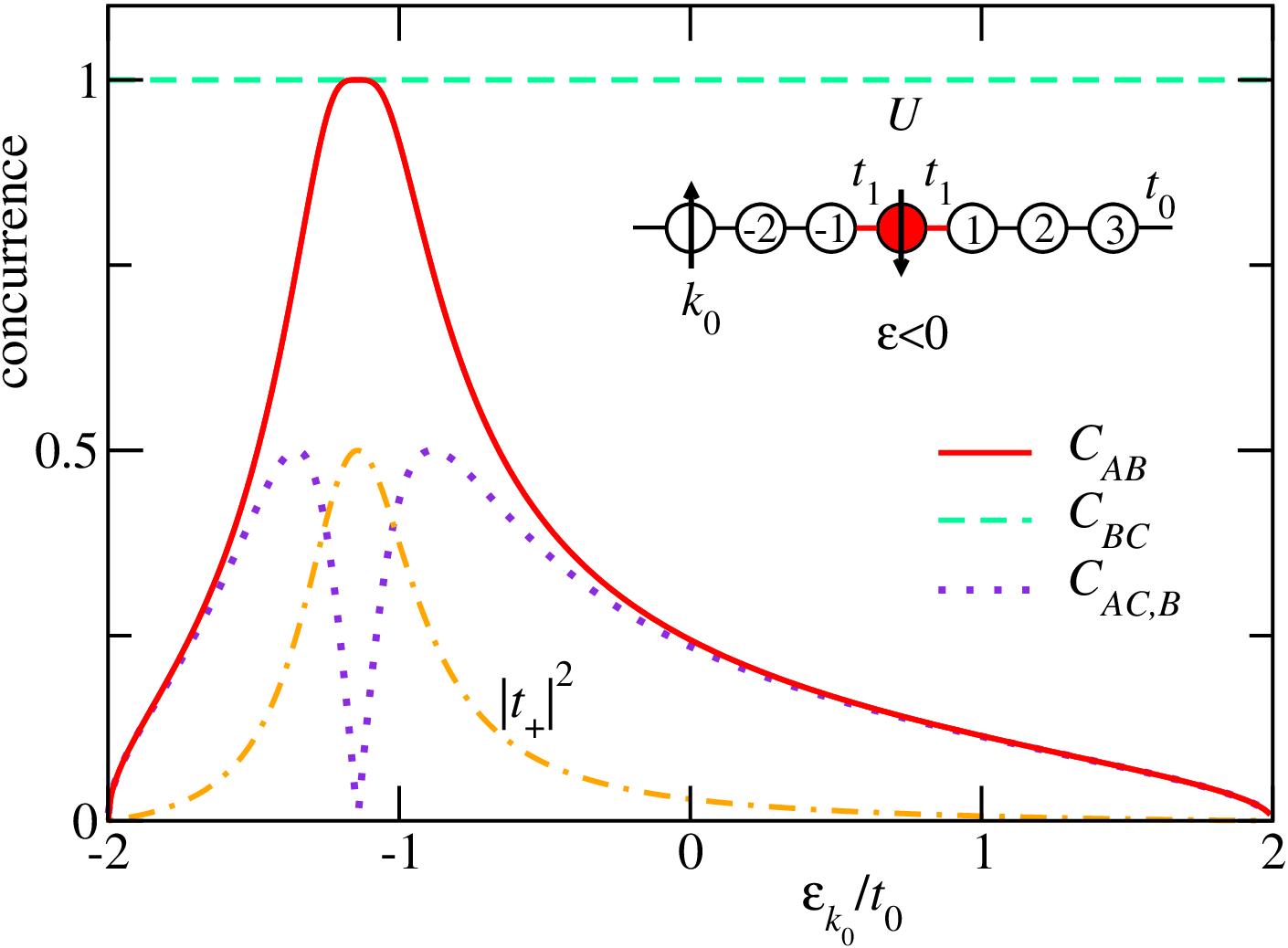}\end{center}

\caption{\label{cap:Fig3}(Color online) Concurrence corresponding to various
domains for infinite-$U$ Anderson model with $\epsilon+U=-t_{0}$,
$t_{1}=t_{0}/4$ and $\Delta k\rightarrow0$. Dashed dotted line represents
the singlet transmission probability $|t_{+}|^{2}$.}
\end{figure}

In the large-$U$ regime, $U,-\epsilon\gg t_{0}$, the static electron
is strongly localised, $b_{i}\sim\delta_{i0}$. Electrons in the triplet
channel are reflected, $r_{-}=-\frac{1}{\sqrt{2}}$, $t_{-}=0$, while
singlet scattering amplitudes exhibit 'charge transfer' resonance:
$t_{+}=\frac{1}{\sqrt{2}}+r_{+}=\frac{\imath}{\sqrt{2}}\Gamma_{k}/(\epsilon_{k}-\omega_{0}+\imath\Gamma_{k})$
with $\epsilon_{k}=-2t_{0}\cos k$, $\omega_{0}=(\epsilon+U)/(1-2t_{1}^{2}/t_{0}^2)$
and $\Gamma_{k}=2t_{1}^{2}(4t_{0}^{2}-\omega_{k}^{2})^{1/2}/(t_{0}^{2}-2t_{1}^{2})$
\cite{ramsak06}. 'Transmitted' concurrence is due to the missing
triplet amplitude, trivially, $C_{BC}\equiv1$. Reflected electrons
are completely entangled at the singlet resonance energy but 'total'
concurrence $C_{AC,B}=0$ there, as shown in Fig.~(\ref{cap:Fig3}).

The main result of this work is the closed form formulae of Wootters
entanglement measure defined for two delocalised electrons. The proposed
approach enables simple analysis of entanglement for a variety of
realistic problems, from scattering of flying and static qubits represented
as wave packets with finite energy resolution, to time evolution of
static qubits due to electron-electron interaction or due to externally
applied fields. Further application to systems described with mixed
states or with more than two electrons is possible, however, an appropriate
definition of entanglement valid also for systems with non-negliglible
doubly occupancy, remains open.

We thank L. Amico, C.W.J. Beenakker, G. Falci, A. Fubini, and T.P. Spiller
for helpful discussions. AR and IS acknowledge support from the Slovenian
Research Agency under contract Pl-0044, and JHJ support from the UK
Ministry of Defence.


\begin{thebibliography}{10}
\bibitem{coulomb} This symmetry is conserved, e.g., in various electron-electron
processes with Coulomb interaction.
\end{thebibliography}
\end{document}